\documentclass[conference]{IEEEtran}
\IEEEoverridecommandlockouts
\usepackage[noadjust]{cite}
\usepackage{amsmath,amssymb,amsfonts}
\usepackage{algorithmic}
\usepackage{graphicx}
\usepackage{textcomp}
\usepackage{xcolor}
\usepackage{subfigure}
\usepackage{url}
\def\BibTeX{{\rm B\kern-.05em{\sc i\kern-.025em b}\kern-.08em
    T\kern-.1667em\lower.7ex\hbox{E}\kern-.125emX}}

\usepackage[absolute]{textpos}

\begin{document}

\begin{textblock}{13}(1.5,0.2)
{\footnotesize \noindent S. Pulipati, V. Ariyarathna, A. Dhananjay, M. E. Eltayeb, M. Mezzavilla, J. M. Jornet, S. Mandal, S. Bhardwaj and A. Madanayake, ``Xilinx RF-SoC-based Digital Multi-Beam Array Processors for 28/60~GHz Wireless Testbeds," \textit{2020 Moratuwa Engineering Research Conference (MERCon)}, Srilanka, Jul 2020, pp. 1-6.}
\end{textblock}

\title{Xilinx RF-SoC-based Digital Multi-Beam Array Processors for 28/60~GHz Wireless Testbeds} 
%{\footnotesize \textsuperscript{~}~}
\author{\IEEEauthorblockN{Sravan Pulipati\IEEEauthorrefmark{1}, Viduneth Ariyarathna\IEEEauthorrefmark{2}, Aditya Dhananjay\IEEEauthorrefmark{3}, Mohammed E. Eltayeb\IEEEauthorrefmark{4}, Marco Mezzavilla\IEEEauthorrefmark{3}, \\ Josep M. Jornet\IEEEauthorrefmark{2}, Soumyajit Mandal\IEEEauthorrefmark{5}, Shubhendu Bhardwaj\IEEEauthorrefmark{1} and Arjuna Madanayake\IEEEauthorrefmark{1}}
\vspace{1ex}
\IEEEauthorblockA{\IEEEauthorrefmark{1}Electrical and Computer Engineering, Florida International University, Miami, USA. Email: amadanay@fiu.edu}
\IEEEauthorblockA{\IEEEauthorrefmark{2}Electrical and Computer Engineering, Northeastern University, Boston, USA. Email: j.jornet@northeastern.edu}
\IEEEauthorblockA{\IEEEauthorrefmark{3} NYU and Pi Radio Inc., Brooklyn, NY, USA. Email: aditya@courant.nyu.edu}
\IEEEauthorblockA{\IEEEauthorrefmark{4}Electrical and Electronic Engineering, California State University, Sacramento, USA. Email:mohammed.eltayeb@csus.edu}
\IEEEauthorblockA{\IEEEauthorrefmark{5}Electrical and Computer Engineering, University of Florida, Gainesville, USA. Email: soumyajit@ece.ufl.edu}}

\makeatletter
\def\ps@IEEEtitlepagestyle{%
  \def\@oddfoot{\mycopyrightnotice}%
  \def\@evenfoot{}%
}
\def\mycopyrightnotice{%
  {\footnotesize {978-1-7281-9975-7/20/\$31.00~\copyright 2020 IEEE}\hfill}% <--- Change here
  \gdef\mycopyrightnotice{}% just in case
}

%\IEEEoverridecommandlockouts \IEEEpubid{\makebox [\columnwidth]{\textbf{978-1-7281-9975-7/20/\$31.00~\copyright 2020 IEEE} \hfill} \hspace{\columnsep}\makebox[\columnwidth]{}}
\vskip -2ex
\maketitle

\begin{abstract}
Emerging wireless applications such as 5G cellular, large intelligent surfaces (LIS), and holographic massive MIMO require antenna array processing at mm-wave frequencies with large numbers of independent digital transceivers. This paper summarizes the authors' recent progress on the design and testing of 28~GHz and 60~GHz fully-digital array processing platforms based on wideband reconfigurable FPGA-based software-defined radios (SDRs). The digital baseband and microwave interfacing aspects of the SDRs are implemented on single-chip RF system-on-chip (RF-SoC) processors from Xilinx. Two versions of the RF-SoC technology (ZCU-111 and ZCU-1275) were used to implement fully-digital real-time array processors at \textcolor{black}{28~GHz (realizing 4 parallel beams with 0.8~GHz bandwidth per beam) and 60~GHz (realizing 4 parallel beams with 1.8~GHz bandwidth per beam)}. Dielectric lenslet arrays fed by a digital phased-array feed (PAF) located on the focal plane are proposed for further increasing antenna array gain.
\end{abstract}


\section{Introduction}
% go up to about 300 GHz with much interest in
The mm-wave frequency range (24-300~GHz) is of increasing interest since they can support large-bandwidth wireless communication channels. Beginning with 24~GHz for continental Europe (and 28~GHz for North America), the available mm-wave spectrum includes i) a license-free band at 60 GHz that is useful for commercial indoor communications systems; and ii) bands at 120-140~GHz, 200-240~GHz, and $>300$~GHz that can support both indoor and outdoor communications. However, the high path losses encountered at higher carrier frequencies requires corresponding increases in either transmit power or directional gain to meet a given link budget. The former is not efficient since this can cause interference to neighboring receivers, so increased antenna gain at both the transmitter and receiver has become the primary method for meeting link budget requirements at mm-wave bands~\cite{Rappaport19,Han15}. 

The presence of high-gain transmit antennas causes mm-wave propagation to become highly-directional. The channels for mm-wave wireless systems employing such antennas are therefore ``ray-like'' because of the directional gain imposed by the transmit beamforming~\cite{rap,rap1}. Although transmit beamforming focuses the available mm-wave radio-frequency (RF) power towards an intended direction, such ray-like propagation makes the communication link susceptible to occlusions when an obstruction appears between the transmitter and the receiver. The presence of occlusions requires exploitation of multiple reflected paths in addition to the primary line of sight (LOS) path, which in turn, requires either multiple RF beams that can be steered when necessary, or a large number of fixed RF beams that can be selected as part of a codebook approach to adaptive beamforming~\cite{Wang09,Chen11}. 

Analog phase-shifter based beamforming has been the predominant approach for achieving highly-directional propagation at mm-wave frequencies~\cite{Venkateswaran10}. However, when the system capacity scales, the number of simultaneous connections across multiple receiver-transmitter pairs also increases. Due to the large number of simultaneous beams, it may no longer be practical to implement a dedicated analog phased-shifter for each beam. Instead, such scaled high-capacity systems can employ either low-complexity optics-inspired methods such as the use of dielectric lenses~\cite{Cho18,Zeng16}, or dense aperture arrays that increase RF hardware complexity. In the latter case, each antenna requires a dedicated transceiver for fully-digital multi-beam beamforming in both receive and transmit modes.

The two approaches described above use i) lens antennas with focal plane array (FPA) feeds, and ii) aperture antenna arrays, respectively; both these must be interfaced to a number of RF transceivers and data converters connected to a high-performance digital signal processing (DSP) engine. In this paper, we propose to meet such challenging hardware and DSP requirements by using commercially-available field programmable gate array (FPGA) devices known as RF System-on-Chips (SoCs). These processors integrate high-speed analog-to-digital converters (ADCs) and digital-to-analog converters (DACs) with an FPGA fabric, hard processor cores, and other peripherals on the same die. 
\IEEEpubidadjcol
\begin{figure*}[t]
\centering
\includegraphics [width=0.9\textwidth] {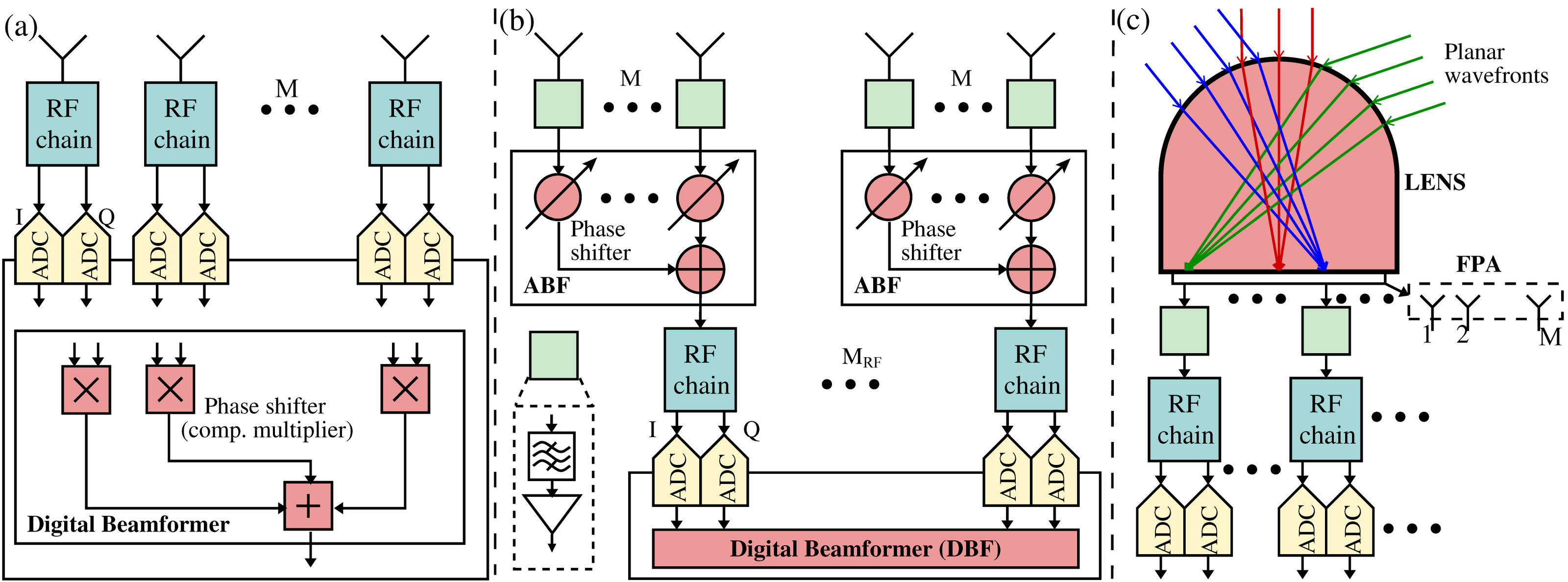}
\vskip -2ex
\caption{Different approaches of beamforming at the receiver side. In all the approaches red blocks indicate the beamformer location. (a) Fully digital beamformer implemented entirely in a digital hardware platform. (b) Hybrid (analog + digital) beamformer where phase shifters are used for Level-1 analog beamforming on a sub-array and level-2 beamforming is performed on the digital hardware platform. (c) Lens + focal planar array (FPA)-based multi-beam beamforming. Lens-based hybrid architectures replace the network of phase-shifters with a lens in order to reduce the complexity and power consumption~\cite{Cho18}.}
\label{block_dia}\vskip -3ex
\end{figure*}
\section{Platforms for Experimental Verification of 5G/6G Communication Theory and Algorithms}
% \section {MIMO Architectures for mmWave Communications}
Communication in the mm-wave bands is a solution that furnishes abundant spectrum for both 5G as well as emerging 6G networks \cite{COSMOS_testbed}. 
To combat the high path-loss at these frequencies, beamforming will be needed at both the transmitter and the receivers. As a consequence of relatively short RF carrier wavelengths in the mm-wave bands (3-30 mm, typical),  large antenna arrays can be packed into small form factors~\cite{Abbaspour03,Biglarbegian11}, thus enabling the deployment of large arrays needed for beamforming. Because mm-wave antenna elements become smaller in area as the square of the wavelength, higher mm-wave frequencies enable more antennas to be packed into a given area of real estate (say, a 2~cm $\times$ 2~cm square on the back of a phone). In other words, the available gain from a given physical aperture increases with increasing frequency. 

\subsection {Digital and Hybrid Beamformers}
The cost and power consumption of mixed signal components in the mm-wave transceivers, however, makes it sometimes unrealistic to assign a dedicated RF chain for each antenna when the number of elements become very large~\cite{Madhow09,Ayach14,Heath16}.  For this reason, several approaches have been considered for the design of low-complexity mm-wave antenna architectures and transmission schemes. Analog beamforming solutions based on analog antenna architectures were proposed in~\cite{Wang09,Hur13}. These architectures consist of a single RF chain and a network of phase shifters connected to each antenna. Despite the reduced complexity, communication schemes based on analog architectures are limited by the quantized phase shifters and single-stream transmission. To overcome these limitations, hybrid analog/digital antenna architectures~\cite{Venkateswaran10,Tayeb15} enable the beamformer to be split between the RF analog and baseband digital domains at the expense of additional RF chains. Digital processing provides even more degrees of freedom and enables multi-beam transmission. Both analog and hybrid antenna architectures for mm-wave systems are mainly based on phased arrays, with the exception of some lens-based architectures~\cite{Brady13,Sayeed13,Joumayly} that replace the phase-shifters with a lens to reduce complexity and power consumption. Fully-digital mm-wave beamforming (also known as digital aperture array beamforming), however, allows the maximum degrees of freedom and the best possible flexibility for signal processing algorithms. However, it requires a dedicated RF transceiver and ADC/DAC pair for each element in the array.

\subsection{Wireless Theory Background}
Fig.~\ref{block_dia} shows a simplified block diagram of a mm-wave hybrid MIMO architecture based on phase-shifters; the precoding and combining operations are split between the analog and the digital baseband domains~\cite{Alkhateeb13,Alkhateeb14}. Observe that when the number of RF chains is reduced to one, this hybrid architecture converges to the conventional analog architecture. Both the transmitter, with $N$ antennas and $N_{RF}\ll N$ RF chains, and the receiver, with $M$ antennas  and $M_{RF}\ll M$ RF chains, communicate over $N_\text{S}\leq \min (N_\text{RF}, M_\text{RF})$ data streams~\cite{Alkhateeb14}.  To transmit the $N_\text{S}$ symbols, the transmitter applies a small $N_\text{RF} \times N_\text{S}$ digital precoder $\mathbf{F}_\text{BB}$ followed by a large $N \times N_\text{RF}$ analog precoder (antenna phase-shifts), $\mathbf{F}_\text{RF}$. The resulting transmitted symbol becomes $\mathbf{x} = \mathbf{F}_\text{RF}\mathbf{F}_\text{BB} \mathbf{a}$, where \textcolor{black}{ $\mathbf{a} = [a_1, a_2, ..., a_\text{N$_\text{s}$}]^\mathrm{T}$ } is the normalized transmitted symbol vector.  At the receiver, the received signals on all antenna elements are combined to obtain 
\begin{eqnarray}\label{hr1}
  \mathbf{y} =   \mathbf{W}^H_\text{BB}\mathbf{W}^H_\text{RF}\mathbf{H}\mathbf{x} + \mathbf{W}^H_\text{BB}\mathbf{W}^H_\text{RF} \mathbf{n},
 \end{eqnarray}
where $ \mathbf{y} \in \mathcal{C}^{N_\text{S}\times 1}$,  $\mathbf{W}_\text{RF}$ is an $M \times M_{\text{RF}}$ RF combining matrix,  $\mathbf{W}_\text{BB}$  is an $M_{\text{RF}} \times N_{\text{S}}$ digital combining matrix, $\mathbf{H}$ is the $M \times N$ matrix that represents the mm-wave channel between the transmitter and the receiver, and $\mathbf{n}$ is the additive noise vector at the receiver. \textcolor{black}{The superscript $H$ in (\ref{hr1}) denotes the Hermitian (conjugate transpose) operator.} One may infer from (\ref{hr1}) that increasing the number of RF chains permits additional data streams and hence higher capacity.  However, this is not the case: the limited number of paths (which may be spatially correlated) in mm-wave channels $\mathbf{H}$ (see (\ref{channelkk})) diminishes the multiplexing gains achievable with large arrays.

%\begin{figure*}[t]
%\centering
%\includegraphics [width=4.5in] {block_dia3.png}
%\caption{Block diagram of a mm-wave hybrid MIMO architecture with $N$ antennas  and $N_{RF}\!\!<<\!\!N$ RF chains at the transmitter and $M$ antennas  and $M_{RF}\!\!<<\!\!M$ RF chains at the receiver. Lens based hybrid architectures replace the network of phase-shifters with a lens in order to reduce the complexity and power consumption \cite{Cho18}.}
%\label{block_dia}
%\end{figure*}

Unlike lower frequency systems, the high propagation path-loss at mm-wave frequencies leads to limited scattering. Further,  mutual coupling between antenna elements as a result of the tightly packed arrays  may lead to further performance degradation. Therefore, classical lower frequency MIMO channel models are not suited to model mm-wave channels. Instead, we use the well-known extended Saleh-Valenzuela channel model~\cite{Alkhateeb14,rap,rap1}. In this case, the discrete narrowband channel $\mathbf{H}$ can be written as
	\begin{eqnarray}\label{channelkk}
	\mathbf{H} = \sqrt{\frac{N_{\text{}}M_{\text{}}}  {CL}} \sum_{c=1}^C \sum_{\ell=1}^L \alpha_{c,\ell} {\mathbf{a}}_\text{r}(\theta^\text{AOA}_{c,\ell},\phi^\text{AOA}_{c,\ell}){\mathbf{a}}_\text{t}^H(\theta^\text{AOD}_{c,\ell},\phi^\text{AOD}_{c,\ell}),
	\end{eqnarray}
where $C$ is the total number of scattering clusters, $L$ is the number of propagation subpaths per cluster, $\alpha_{c,\ell}$ is the complex gain of the $c,\ell$th subpath,  $\theta_{c,\ell} \in [0, 2\pi]$  $\phi_{c,\ell} \in [0, 2\pi]$ are the $c,\ell$-th subpath's azimuth and elevation angles of departure or arrivals (AoD/AoA). The vectors  $\mathbf{a}_\text{t}(\theta^\text{AOD}_{c,\ell},\phi^\text{AOD}_{c,\ell})$ and 
$\mathbf{a}_\text{r}(\theta^\text{AOA}_{c,\ell},\phi^\text{AOA}_{c,\ell})$  denote the $N \times 1$ transmitter and $M \times 1$ receiver array response vectors. Note that the limited number of channel paths, possibly correlated, result in diversity and multiplexing gain loss in mm-wave systems. Nonetheless, this channel property can be exploited to further reduce system complexity by minimizing the number of RF chains. 

\subsection{Synopsis of Hardware Development Efforts}
The rest of the paper is organized as follows. In the next section, we describe recent progress in multi-beam beamforming using a dielectric lens antenna fed by a 4-element FPA located on the focal region of the lens. The lens is an analog passive device that uses refraction to compute a spatial Fourier transform in the mm-wave band of interest. It was designed for 28~GHz operation in mind, and the FPA is connected to 4 dedicated direct-conversion receivers. In the subsequent section, we describe our recent progress with fully-digital multi-beam beamforming at 60~GHz. The system, which is under development at Pi-Radio (Brooklyn, NY)\footnote{Pi-Radio is a startup supported by the New York State Center for Advanced Technologies in Telecommunications (CATT) at  New York University.}, consists of 4-element transmit and receive arrays. Each array can also operate as an FPA for 60~GHz lens antennas when required. Preliminary experimental results are reported for both systems.

\section{Multi-beam 28~GHz Beamforming using a Lens Antenna}
\begin{figure}[!t]
\begin{center}
\includegraphics[width=0.48\textwidth]{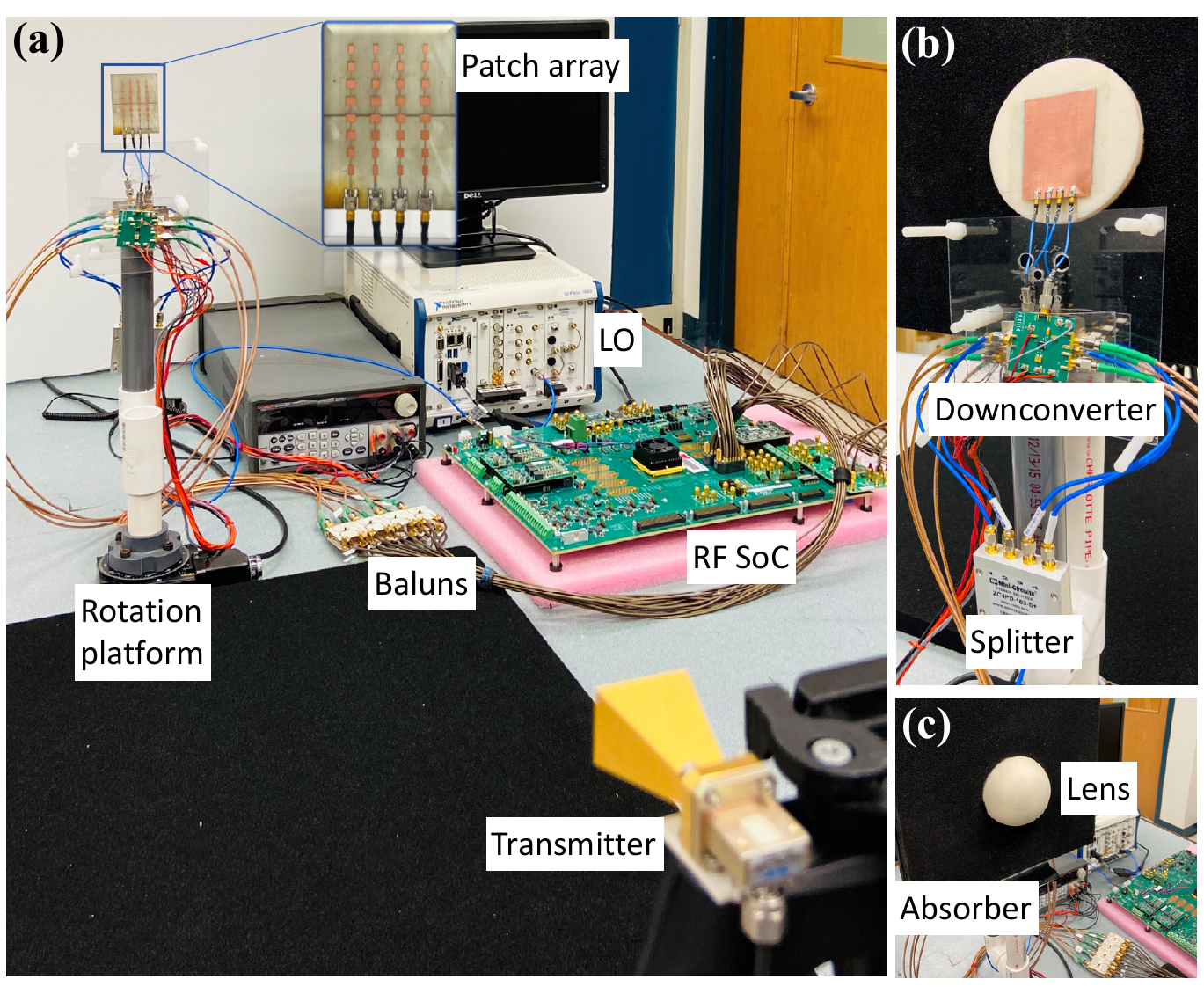}
\vskip -2.5ex
\caption{(a) Experimental setup containing the 28~GHz digital array receiver with digital back-end RF-SoC system (Xilinx ZCU-1275)  (b) Receiver chain for 4-element FPA with array taped to the base of the lens. A Mini-Circuits four-way power splitter distributes the 14~GHz local oscillator (LO) to all mixers. Each transceiver contains a frequency doubler that generates 28~GHz LOs from the 14~GHz distribution network. (c) Front view of the lens + FPA.\label{fig:28GHz_array_ov}}
\end{center}
\vskip -4ex
\end{figure}

Fig.~\ref{fig:28GHz_array_ov} shows the 4-element lens + FPA-based digital receiver at 28 GHz. The digital back-end employs a Xilinx ZCU-1275 RF-SoC to generate four 28~GHz RF receive-mode beams with a bandwidth of 0.85~GHz per beam. The array design, specifications and the measurements are discussed next.

\subsection{28 GHz Digital Array Receiver Specifications}
A hemispherical lens of radius $r=5$~cm and cylindrical base of length 5.7~cm was designed and 3-D printed using acrylonitrile butadiene styrene (ABS). The lens dimensions are calculated to yield a directivity of 29.3~dBi for $\lambda = 10.7$~mm at 28~GHz.  The FPA consists of a 28~GHz 4-element patch antenna array supported by 4 down-converters. Each antenna element in the array is itself built as an 8-element series-fed array to provide additional gain in the elevation plane \cite{COMCAS2019}. \textcolor{black}{The LENS and antenna arrays are fabricated with an accuracy of 0.1 mm}. For the 28~GHz receivers, we use a downconverter chip (AD-HMC1065LP4E) to mix down the RF signal to baseband, which supports up to 1~GHz I-Q channel bandwidth. The I-Q down-converted signals are passed through balanced-unbalanced RF transformers, i.e., baluns (TCM2-33WX+, Mini-Circuits) so as to interface with the ADCs of the Xilinx RF-SoC, which have differential inputs.  %This is a requirement of the particular RF-SoC platform.

The digital back-end was implemented using the Xilinx ZCU-1275~\cite{1275UserGuide} development platform, which incorporates a Zynq UltraScale+ XCZU29DR RF-SoC~\cite{RFSoC}. This chip contains 16 parallel 12-bit ADCs, each capable of sampling rates up to 2~GSps. The RF-SoC also contains 16 parallel 14-bit RF DACs, each capable of 6.4~GSps. \textcolor{black}{Calibration of the RF front-ends was performed digitally. Initially we measure the gain and phase mismatches between chains using a reference 28 GHz signal, and then employing digital complex multipliers to compensate those mismatches.} Comprehensive details of receiver design and specifications along with digital back-end details can be found in \cite{COMCAS2019,IMS2020}. 

\subsection{Multi-Beam Measurements Setup for 28 GHz Lens}
The experimental setup used to measure the RF beams is shown in Fig.~\ref{fig:28GHz_array_ov}. A 28~GHz horn antenna is used as a transmitter, and the LO frequency is set to 27.9~GHz. Eight ADC channels of the Xilinx RF-SoC are configured to sample the four I-Q signals at 1966.08~MHz. To emulate the direction of arrival for beampattern measurements, the array setup is rotated along its axis keeping the transmitter fixed. ADC samples from each channel are passed onto a digital integrator block to compute the gain for that particular angle.
\begin{figure*}[!t]
\begin{center}
\includegraphics[width=\textwidth]{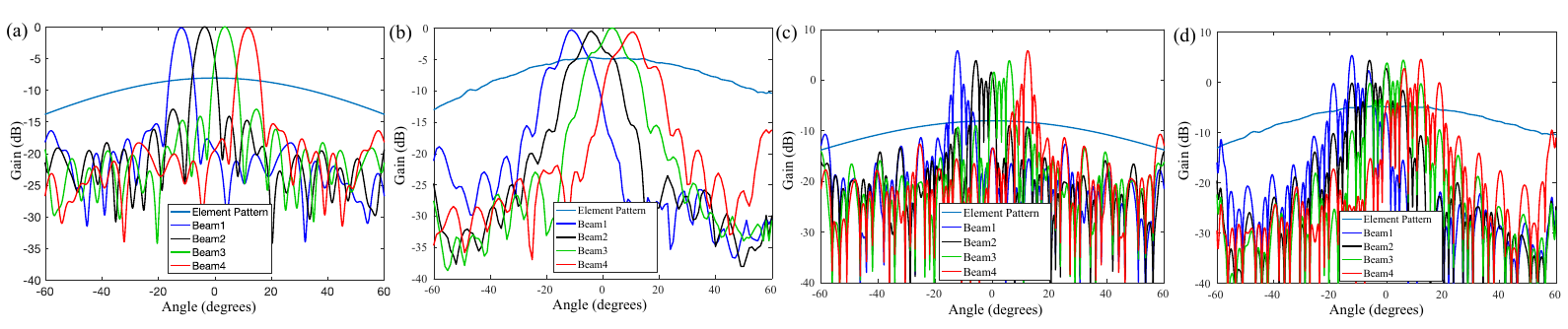}
\end{center}
\vskip -4ex
\caption{(a) CST-simulated beam patterns. The elemental subarray gain pattern (normalized by the maximum of beam pattern) is shown for comparison. (b) Measured beam patterns and corresponding normalized elemental subarray pattern~\cite{IMS2020}. (c) Simulated beam patterns from lens + FPA setup for a lenslet array. (d) Simulated beam patterns obtained from cascaded lens + FPA setup using measured lens beam data. \label{fig:beamswithelementpatterns}}
\end{figure*}

\begin{figure}[!h]
\vskip -2ex
\begin{center}
\includegraphics[width=0.3\textwidth]{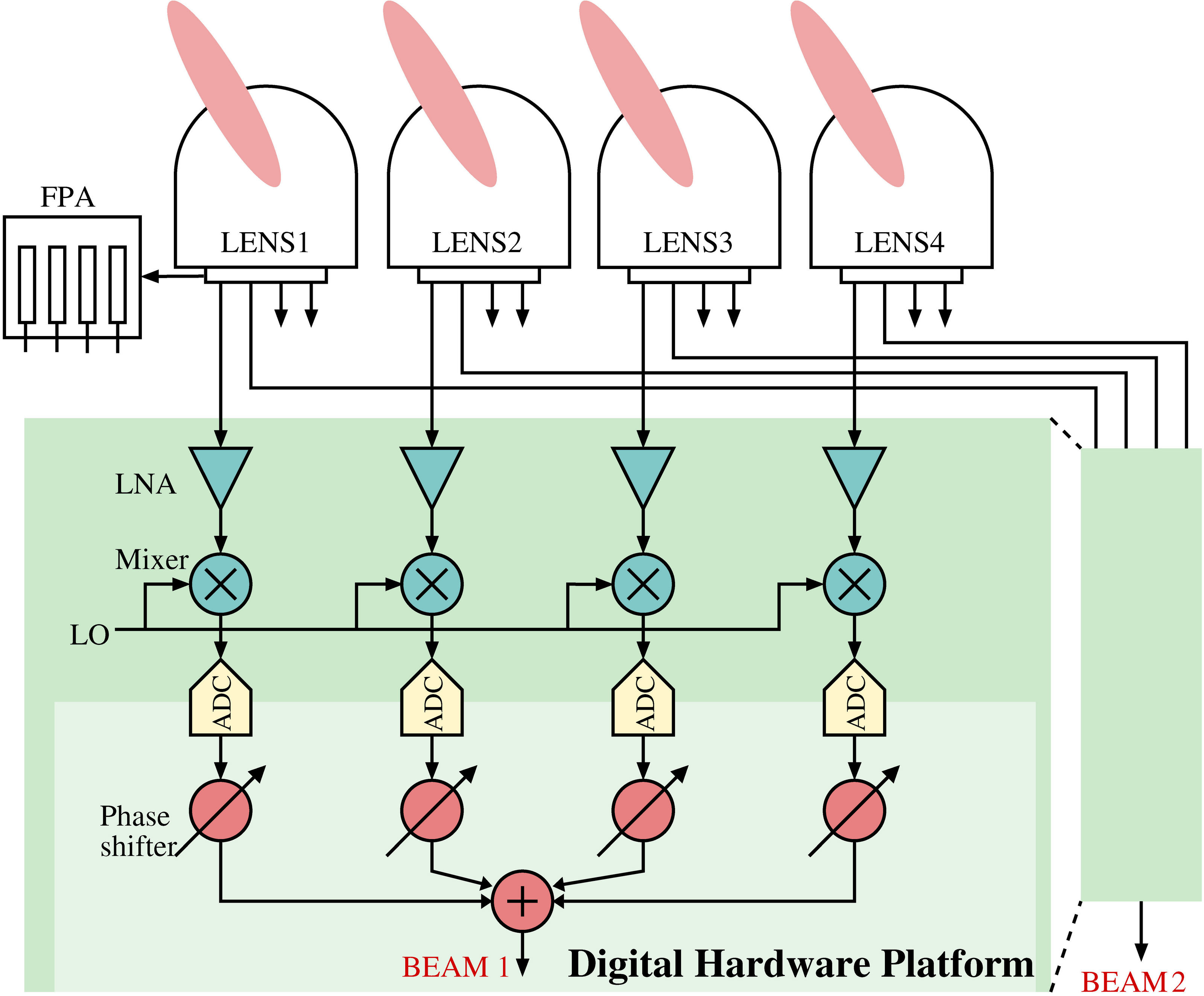}
\end{center}
\vskip -3ex
\caption{System overview of the cascaded lens + FPA multibeam beamformer used to generate sharper beams. \label{fig:lenslet}}
\vskip -3ex
\end{figure}
\subsection{Measured 28 GHz Lens + FPA Digital Beams}
Fig.~\ref{fig:beamswithelementpatterns}(a) shows CST-simulated beam patterns of the lens + FPA configuration. The elemental pattern without the lens (shown in cyan) is included to show the effect of lens gain.  The measured receive-mode RF beams obtained from the RF-SoC are shown in Fig.~\ref{fig:beamswithelementpatterns}(b). The beams point at the correct angles and are in agreement with the simulations; however, the gain is smaller than expected. We expect this mismatch to be due to i) material losses in the ABS material, and ii) a slight offset in the position of the FPA, which can cause major changes in the gain. These effects will be studied in future work.

\subsection{Simulated 28 GHz Beams using Lenslets}
The beams obtained from the original setup can be further narrowed down by cascading several such array structures and feeding to a phased array digital beamformer. A system overview of such lenslet approach is shown in Fig.~\ref{fig:lenslet}.  A MATLAB simulation is set up to illustrate this effect using measured data from a single lens. The setup assumes 4 lenses + FPA (i.e., the current setup) placed next to each other, such that the distance between two corresponding antenna elements is equal to the lens diameter, i.e., 10~cm. To compute the beampatterns for this setup, the measured beampattern for one antenna was multiplied by the array factor of a four-element phased array beamformer. Figs.~\ref{fig:beamswithelementpatterns} (c) and (d) show the simulated and measured beampatterns respectively. The lensets generate narrow beams, which in turn provide additional gain.

\section{Digital 60~GHz Array Processor on ZCU-111 RF SoC}
\begin{figure}[htbp]
\centering
\includegraphics[width=0.70\columnwidth]{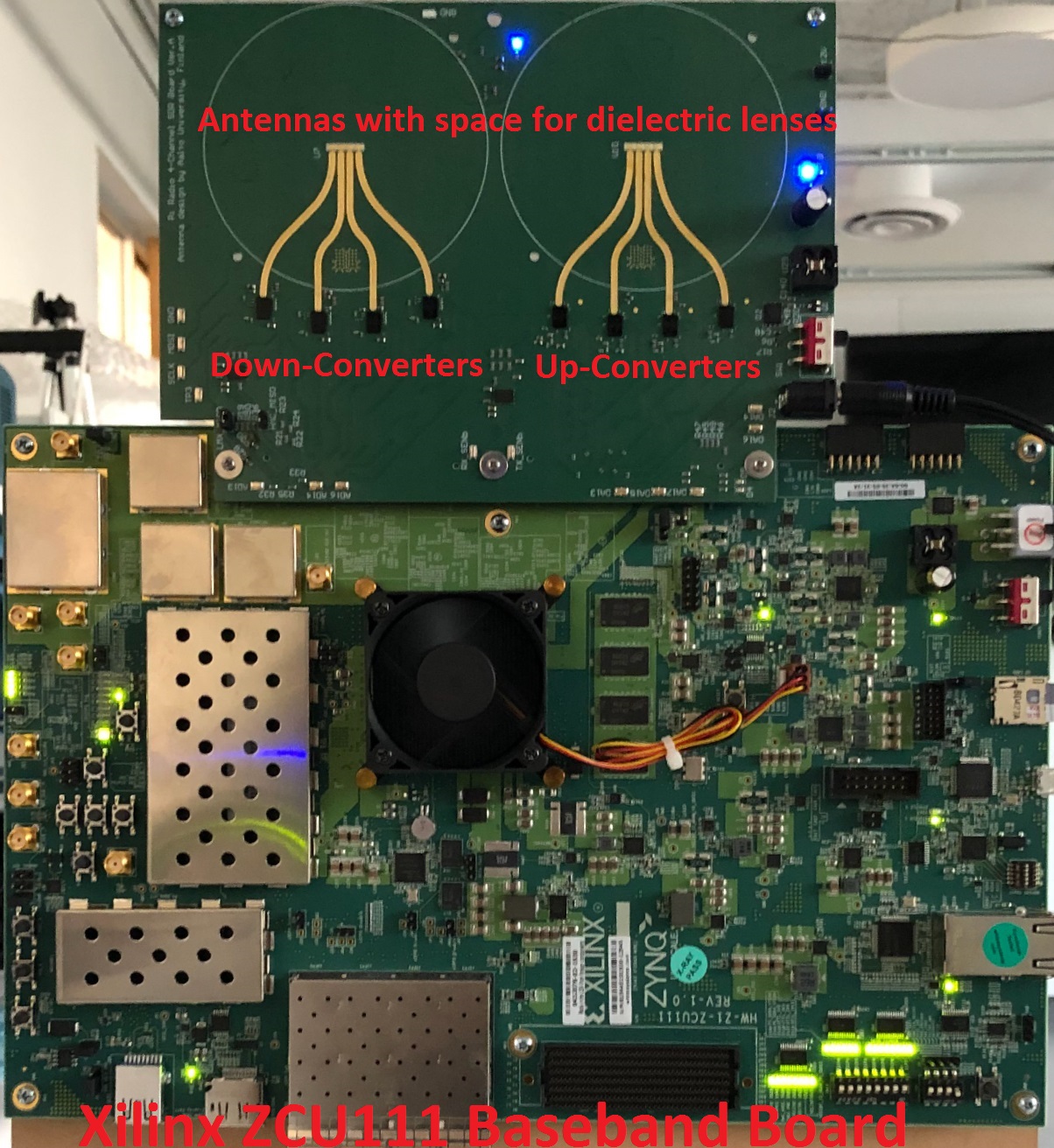}
\vskip -2ex
\caption{Photo of the SDR system featuring the Pi-Radio fully-digital
transceiver board (57-64~GHz) paired with the Xilinx ZCU-111 RF-SoC platform. The transceiver board features mm-wave up- and down-converter chips, along with an LO generation, amplification, and distribution network. The 57-64~GHz $1 \times 4$ patch antenna array was designed at Aalto University (Finland). Observe the circular
{\em keep-out} areas around the antennas; these are to mount dielectric lenses
using the provided flange holes.}
\label{fig:piradio-photo}
\vskip -2ex
\end{figure}

Pi-Radio has developed a 4-channel fully-digital transceiver board operating in the 57-64 GHz band (shown in Fig.~\ref{fig:piradio-photo}). This custom front-end is used with the
Xilinx ZCU111 RF-SoC board; along with a host computer, it forms an SDR system for use by the wireless research community.
\subsection{60 GHz Array Board Hardware}
\begin{figure*}[!ht]
\centering     %%% not \center
\vskip -1ex
%\subfigure{\includegraphics[height=37mm]{images/antenna_testing_aalto}}
%\subfigure{\includegraphics[height=37mm]{images/specan_ssb}}
%\subfigure{\includegraphics[height=37mm]{images/rx_ssb}}
\includegraphics[width=0.75\textwidth]{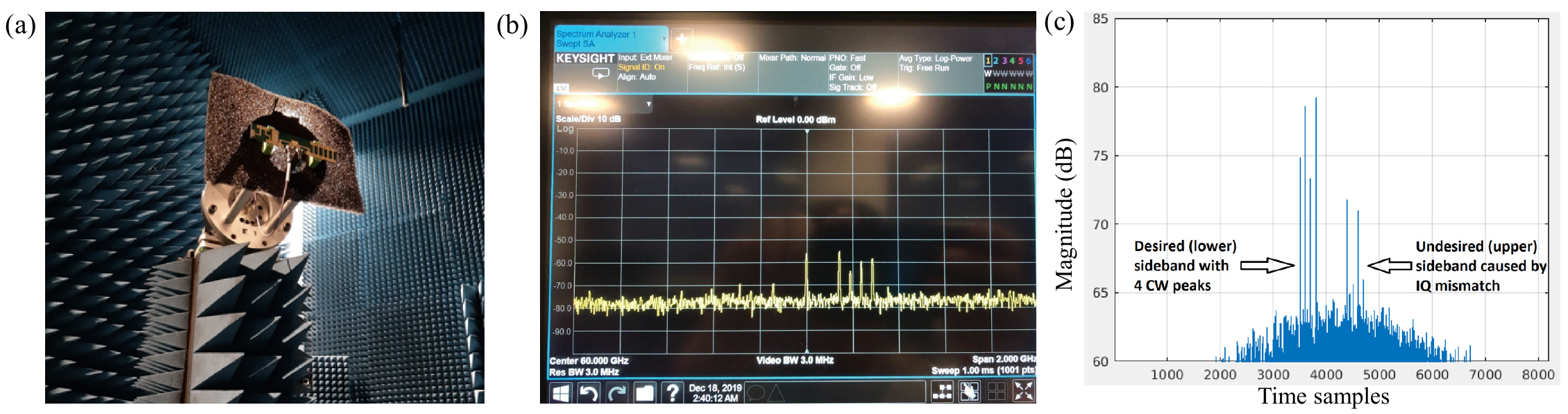}
\vskip -1ex
\begin{scriptsize}
\caption{Left to right: (a) Testing of the antenna array using an
anechoic chamber at Aalto University, Finland. (b) Four CW peaks captured on a spectrum analyzer during testing of the four TX chains. (c) Four CW peaks captured in MATLAB (raw signal, prior to calibration) from one RX chain in a frequency-selective fading environment; note that I/Q calibration will further suppress the unwanted sideband.}
\label{fig:antenna-peaks}
\end{scriptsize}
\vskip -3ex
\end{figure*}

On the Tx side, the digital baseband I-Q samples are generated within the RF-SoC FPGA,
and are converted to analog using the DACs present on the chip. There are
8 DACs (I and Q use separate DACs), leading to operation over four RF channels.
The resulting analog I-Q data is sent to the Pi-Radio transceiver board using the
LPAM/LPAF connector pairs. A bank of four mm-wave
up-converters (HMC6300, Analog Devices) converts these baseband signals to RF. The RF signals are routed to a $1\times4$ patch antenna array that has been designed at Aalto University in
Finland~\cite{aalto-antenna}; a photograph of the antenna measurements being
conducted is shown in Fig.\ref{fig:antenna-peaks}(a). The Rx side is symmetric,
with the RF signals from the antennas being converted to baseband using mm-wave down-converters (HMC6301, Analog Devices). The received baseband analog I-Q streams are sent to
the RF-SoC device using LPAM/LPAF connector pairs, and digitized using the 8 parallel RF ADCs within the RF-SoC.

A fully-digital beamformer requires all the Tx and Rx channels on a node to be phase-locked (time-synchronized) with respect to each other. The RF up- and down-conversion chips utilize an external LO signal that is generated,
amplified, and distributed on the Pi-Radio transceiver board. Within the up- and down-converters,
this LO signal is split into two paths: a) power-divided by 2 using 3-dB splitters, and used to convert between baseband and intermediate frequency (IF); and b) frequency multiplied by 3, and
used to convert between IF and RF. Thus, the RF frequency is $3.5\times$ the external LO frequency. For example, if 60~GHz operation is desired, the external LO frequency is $\frac{60}{3.5}=17.142857$~GHz. This LO signal is generated by an RF frequency synthesizer (LMX2595, Texas Instruments), and the differential outputs are first split into two single-ended signals using an impedance matching circuit. These two paths (one for the Tx chains and one for
the Rx chains) are independently amplified using two RF amplification stages (HMC441 and HMC962, Analog Devices). The amplified outputs are passed through 1:4 Wilkinson power dividers (Knowles Dielectric Labs). The outputs of the two splitters are fed to the up- and down-converters,
thereby ensuring that all the Tx and Rx chains are phase-locked. The boards feature large circular {\em keep-out} areas around the patch antennas, thereby allowing third-party dielectric lens antennas to be mounted using the provided flange holes.

\subsection{Software and DSP Algorithms}
The device drivers for this SDR system are based on the Xilinx
RFSoC reference drivers, which are responsible for the following: a)
configuring the DACs and ADCs; b) moving data between the FPGA and the host;
and c) running Petalinux on the ARM cores of the RFSoC to implement this
functionality. Pi-Radio has augmented these drivers with functionality to
control the transceiver board.  All hardware schematics and
source code are free and open-source, and hosted on GitHub.

\subsection{Test and Measurement}
 While final calibrations and bring-up of the SDR system are
ongoing, we present some early results. Fig.~\ref{fig:antenna-peaks} (b)
demonstrates the operation of the 4 Tx chains. Each chain generates a
different continuous wave (CW) tone, and the transmitted signals are examined using a spectrum
analyzer (N9020B, Keysight) with an external waveguide harmonic mixer (M1971V, Keysight) to extend its frequency range to 75~GHz). Fig.~\ref{fig:antenna-peaks} (c) demonstrates the operation of the 4 Rx chains. The signals transmitted from one SDR unit (described previously) are captured using the Rx chains of a second SDR unit. The raw signals (prior to being digitally calibrated) from one such Rx chain are plotted in MATLAB. Observe the frequency-selective fading in the wireless channel. I-Q imbalance calibration will further suppress the unwanted sideband.

\begin{figure}[t]
\begin{center}
\includegraphics [width=0.70\columnwidth] {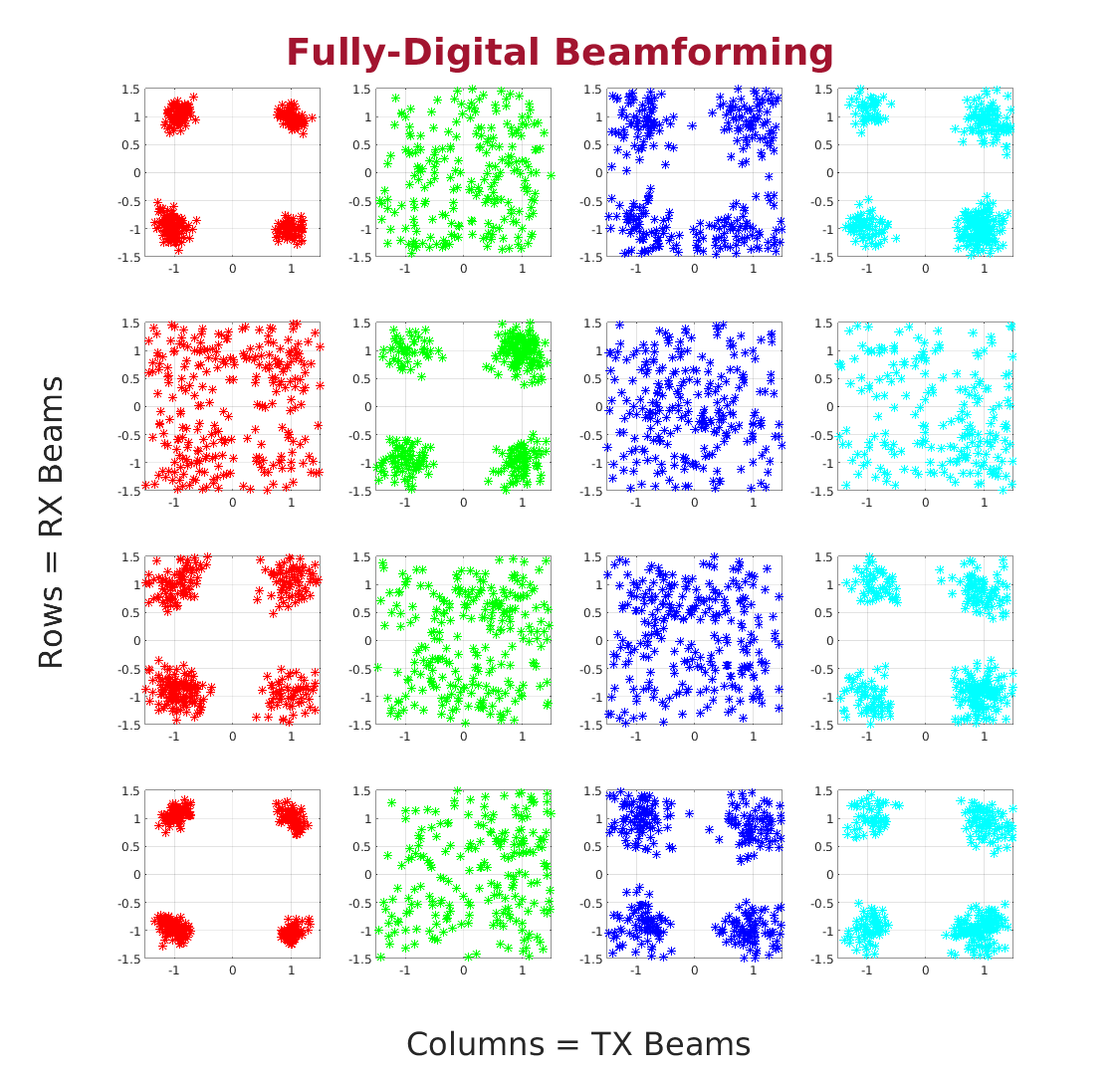}

\vskip -2ex
\caption{A fully-digital beamformer can \textit{look} in all directions
simultaneously. Constellation plots show the received quadrature phase shift keyed (QPSK) symbols at nominal 60~GHz carrier frequency, when decoded in
each of four Rx directions, with each Rx direction decoding data corresponding to four Tx directions. Observe that at least one Rx direction can decode each of the Tx streams. Such $O(1)$ synchronization is a huge improvement over the $O(n^2)$ synchronization needed in analog beamforming (phased-array) systems. \label{fig:constellation2}}
\end{center}
\label{fig:constellation}
\end{figure}

\subsection{Data Link Demonstration}
Two 60~GHz SDR nodes are configured to create a mm-wave data link, with Node~1 as the
transmitter and Node~2 as the receiver. Node~1 transmits four independent streams
of data, wherein each stream is beam-formed in a different direction, and
operates over a different sub-channel in frequency. Node~2 digitally beam-forms in 4
directions simultaneously (by applying different beamforming vectors to the same received time-domain samples), and for each Rx direction, attempts to
synchronize and decode each of the 4 data streams. The results are plotted in
Fig.~\ref{fig:constellation2}. Each row corresponds to a different Rx direction,
and each column to a different Tx stream / direction. Observe that
for each transmitted data stream, there exists at least one Rx direction that
can successfully decode the data. This illustrates a key feature of
fully-digital beamformers: \textit{unlike an analog beamformer (phased array) that can look in only one direction at a time, a digital beamformer can ``look in all directions simultaneously"}. If each node can use $n$ directions, then an analog beamformer would need $O(n^2)$ time to synchronize. By contrast, a digital beamformer can synchronize and decode multiple data streams in $O(1)$ time. This capability is highly significant from a systems perspective.

\subsection{Future Extensions to Sub-THz Bands}
{While the total bandwidth at mm-waves is much larger than that of sub-6~GHz systems, even more bandwidth is needed to support faster (approaching 1~Terabit-per-second or Tbps) and higher node densities,  towards billions of interconnected devices in the wireless internet of things (IoT). This bandwidth can be accessed by moving to even higher frequency bands. There is a growing interest in sub-THz (100-300 GHz) and true THz (0.3-3.0 THz) frequencies, where the channel supports consecutive bandwidths of tens to hundreds of GHz~\cite{akyildiz2014terahertz,sen2019experimental}, in between absorption lines from water vapor molecules.}  

\section{Conclusions}
Multirate digital array processing hardware including antenna arrays, mm-wave front-ends, high-bandwidth basebands, and Xilinx RF-SoC based DSP systems were proposed for aperture/lens antennas operating at 28 and 60~GHz. Digital basebands were realized using the Xilinx ZCU-111 and ZCU-1275 RF-SoC development boards. Preliminary RF measurements and 4-beam beamforming results for 4-element arrays confirm high real-time baseband bandwidths (0.8~GHz and 1.8~GHz per beam, respectively).  Digital multi-beam beamforming using an FPA-fed lenset array was also demonstrated. Future work includes realization of media access control (MAC) and networking layer algorithms on Xilinx RF-SoCs.


\begin{thebibliography}{00}
\bibitem{Rappaport19}
T. S. Rappaport, Y. Xing, O. Kanhere, S. Ju, A. Madanayake, S. Mandal, A. Alkhateeb, and G. C. Trichopoulo, ``Wireless communications and applications above 100 GHz: Opportunities and challenges for 6G and beyond,'' \emph{IEEE Access}, vol. 7, pp. 78729-57, 2019.
\bibitem{Han15} 
S. Han, I. Chih-Lin, Z. Xu, and C. Rowell, ``Large-scale antenna systems with hybrid analog and digital beamforming for millimeter wave 5G,'' \emph{IEEE Commun. Mag.}, vol. 53, no. 1, pp. 186-194, Jan. 2015.
\bibitem{rap}
M. Akdeniz, Y. Liu, M. Samimi, S. Sun, S. Rangan, T. Rappaport, and E. Erkip, ``Millimeter wave channel modeling and cellular capacity evaluation,'' \emph{IEEE J. on Selected Areas in Commun.,} vol. 32, no. 6, pp. 1164-1179, June 2014.
\bibitem{rap1}
T. Rappaport, Y. Qiao, J. Tamir, J. Murdock, and E. Ben-Dor, ``Cellular broadband millimeter wave propagation and angle of arrival for adaptive beam steering systems,'' in \emph{Radio and Wireless Symposium (RWS),} Santa Clara, CA, Jan. 2012, pp. 151-154.
\bibitem{Wang09}
J. Wang et al., ``Beam codebook based beamforming protocol for multiGbps millimeter-wave WPAN systems,'' \emph{IEEE J. Sel. Areas Commun.}, vol. 27, no. 8, pp. 1390-1399, Oct. 2009.
\bibitem{Chen11} 
L. Chen, Y. Yang, X. Chen, and W. Wang, ``Multi-stage beamforming codebook for 60 GHz WPAN,'' in \emph{Proc. 6th Int. ICST Conf. Commun. Netw.}, Harbin, China, Aug. 2011, pp. 361-365.
\bibitem{Venkateswaran10} 
V. Venkateswaran and A. van der Veen, ``Analog beamforming in MIMO communications with phase shift networks and online channel estimation,'' \emph{IEEE Trans. Signal Process.}, vol. 58, no. 8, pp. 4131-4143, Aug. 2010.
\bibitem{Cho18} 
Y. J. Cho, G. Suk, B. Kim, D. K. Kim, and C. Chae, ``RF lens-embedded antenna array for mmWave MIMO: design and performance,'' \emph{IEEE Communications Magazine}, vol. 56, no. 7, pp. 42-48, July 2018.
\bibitem{Zeng16} 
Y. Zeng and R. Zhang, ``Millimeter wave MIMO with lens antenna array: A new path division multiplexing paradigm,'' \emph{IEEE Trans. Commun.}, vol. 64, no. 4, Feb. 2016, pp. 1557-71.
\bibitem{COSMOS_testbed}
Raychaudhuri, Dipankar, et al. "Challenge: COSMOS: A city-scale programmable testbed for experimentation with advanced wireless." Proc. ACM MobiCom. Vol. 20. 2020.
\bibitem{Abbaspour03}    
A. Abbaspour-Tamijani and K. Sarabandi, ``An affordable millimeter-wave beam-steerable antenna using interleaved planar
subarrays,'' \emph{IEEE Trans. Antennas and Propagation}, vol. 51, no. 9, pp. 2193-2202, 2003.
\bibitem{Biglarbegian11} 
B. Biglarbegian, M. Fakharzadeh, D. Busuioc, M. Nezhad-Ahmadi, and S. Safavi-Naeini, ``Optimized microstrip antenna
arrays for emerging millimeter-wave wireless applications,'' \emph{IEEE Transactions on Antennas and Propagation}, vol. 59, no. 5, pp. 1742-1747, 2011.
\bibitem{Madhow09} 
J. Singh, S. Ponnuru, and U. Madhow, ``Multi-Gigabit communication: The ADC bottleneck,'' in \emph{Proc. IEEE Int. Conf. Ultra-Wideband (ICUWB)}, Vancouver, BC, Canada, Sep. 2009, pp. 22-27.
\bibitem{Ayach14} 
O. El Ayach, S. Rajagopal, S. Abu-Surra, Z. Pi, and R. W. Heath, Jr., ``Spatially sparse precoding in millimeter wave MIMO systems,'' \emph{IEEE Trans. Wireless Commun.}, vol. 13, no. 3, pp. 1499-1513, Mar. 2014.
\bibitem{Heath16} 
R. W. Heath, Jr., N. Gonzalez-Prelcic, S. Rangan, W. Roh, and A. Sayeed, ``An overview of signal processing techniques for millimeter wave MIMO systems,'' \emph{IEEE Journal of Selected Topics in Signal Processing}, vol. 10, no. 3, pp. 436-453, April 2016.
\bibitem{Hur13} 
S. Hur, T. Kim, D. J. Love, J. V. Krogmeier, T. A. Thomas, and A. Ghosh, ``Millimeter wave beamforming for wireless backhaul and access in small cell networks,'' \emph{IEEE Transactions on Communications}, vol. 61, no. 10, pp. 4391-4403, Oct. 2013.
\bibitem{Tayeb15}
M Eltayeb, A. Alkhateeb, R. W. Heath, and T. Al-Naffouri, ``Opportunistic beam training with hybrid analog/digital codebooks for mmWave systems,'' in \emph{IEEE Global Conference on Signal and Information Processing (GlobalSIP)}, Dec. 2015, Orlando, Florida, USA.
\bibitem{Brady13} 
J. Brady, N. Behdad, and A. M. Sayeed, ``Beamspace MIMO for millimeter-wave communications: System architecture, modeling, analysis, and measurements,'' \emph{IEEE Trans. Antennas Propag.}, vol. 61, no. 7, pp. 3814-3827, Jul. 2013.
\bibitem{Sayeed13} 
A. Sayeed and J. Brady, ``Beamspace MIMO for high-dimensional multiuser communication at millimeter-wave frequencies,'' \emph{Proc. IEEE GLOBECOM}, Dec. 2013, pp. 3679-84.
\bibitem{Joumayly} 
M. A. Al-Joumayly and N. Behdad, ``Wideband planar microwave lenses using sub-wavelength spatial phase shifters,'' \emph{IEEE Trans. Antennas Propag.}, vol. 59, no. 12, Aug. 2011, pp. 4542-52.
\bibitem{Alkhateeb13} 
A. Alkhateeb, O. El Ayach, G. Leus, and R. W. Heath, Jr., ``Hybrid precoding for millimeter wave cellular systems with partial channel knowledge,'' in \emph{Inf. Theory Appl. Workshop (ITA)}, Feb. 2013, pp. 1-5.
\bibitem{Alkhateeb14} 
A. Alkhateeb, O. El Ayach, G. Leus, and R. W. Heath, Jr., ``Channel estimation and hybrid precoding for millimeter wave cellular systems,'' \emph{IEEE Journal on Selected Topics in Signal Processing}, vol. 8, no. 5, pp. 831-846, Oct. 2014.
\bibitem{1275UserGuide}
Xilinx, ``{ZCU1275} {Characterization Board User Guide} (v1.0),'' Nov. 2018. [Online]. Available:
\url{https://www.xilinx.com/support/documentation/boards_and_kits/zcu1275/ug1285-zcu1275-char-bd.pdf}
\bibitem{RFSoC}
Xilinx\textsuperscript{\textregistered}, ``Zynq {UltraScale} {RFSoC} {Data Sheet} {DC} and {AC} {Switching Characteristics},'' {Available:
  \url{https://tinyurl.com/yc728o6x}}.
\bibitem{COMCAS2019}
S. Pulipati et al., ``A direct-Conversion digital beamforming array receiver with 800 MHz channel bandwidth at 28 GHz using Xilinx RF SoC,'' \emph{IEEE Intl. Conf. Microwaves, Antennas, Communications and Electronic Systems (COMCAS)}, Tel-Aviv, Israel, 2019, pp. 1-5.
\bibitem{IMS2020}
S. Pulipati, V. Ariyarathna, M. Khan, S. Bhardhwaj, and A. Madanayake, ``Aperture array \& lens+FPA multi-beam digital receivers at 28 GHz on Xilinx ZCU-1275 RF SoC,'' \emph{IEEE IMS 2020}, Los Angeles, CA (to appear).
\bibitem{aalto-antenna}
J. Haarla, V. Semkin, K. Zheng, A. Dhananjay, M. Mezzavilla, J. Ala-Laurinho, and V. Viikari, ``Characterizing 60 GHz patch antenna segments for fully digital transceiver,''  \emph{Proc. EuCAP}, 2020.
\bibitem{akyildiz2014terahertz}
I.~F. Akyildiz, J.~M. Jornet, and C.~Han, ``Terahertz band: Next frontier for wireless communications,'' \emph{Physical Communication}, vol.~12, pp. 16--32, 2014.  
\bibitem{sen2019experimental}
P. Sen and J. M. Jornet, ``Experimental demonstration of ultra-broadband wireless communications at true Terahertz frequencies,'' \emph{Proc. IEEE International Workshop on Signal Processing Advances in Wireless Communications (SPAWC)}, Nice, France, July 2019.
\end{thebibliography}
\end{document}